\documentstyle[aps,prd]{revtex}

\def \ep {\varepsilon}
\def \c {\mbox{curl}\,}
\def \d {\mbox{D}}
\def \div {\mbox{div}\,}

\def \ts {\textstyle}
\def \rd {\displaystyle{\cdot}}

\begin{document}
\tighten

\title{Consistency of dust solutions with div $H=0$}

\author{Roy Maartens\footnote{maartens@sms.port.ac.uk}}

\address{School of Mathematical Studies, Portsmouth University,
Portsmouth PO1 2EG, Britain \\
and Department of Mathematics and Applied Mathematics, University of
Natal, Durban 4041, South Africa}

\author{William M. Lesame\footnote{lesame@ufhcc.ufh.ac.za}}

\address{Department of Applied Mathematics, University of Fort Hare,
Alice 5700, South Africa}

\author{George F.R. Ellis\footnote{ellis@maths.uct.ac.za}}

\address{Department of Mathematics and Applied Mathematics,
University of Cape Town, Cape Town 7700, South Africa}

\date{December 1996}

\maketitle

\begin{abstract}

One of the necessary covariant conditions for gravitational radiation 
is the vanishing of the divergence of
the magnetic Weyl tensor $H_{ab}$, while $H_{ab}$ itself is nonzero. 
We complete a recent analysis by showing that  
in irrotational dust spacetimes, the condition
$\div H=0$ evolves consistently in the 
exact nonlinear theory.

\end{abstract}


Irrotational dust spacetimes, typically considered as models for
the late universe or for gravitational collapse, are covariantly
characterized by the dust four--velocity $\,u^a\,$, energy 
density $\,\rho\,$,
expansion $\,\Theta\,$ and shear $\,\sigma_{ab}\,$, 
and by the free gravitational 
field, described by the electric and magnetic parts of the Weyl 
tensor $\,C_{abcd}\,$:
$$
E_{ab}=C_{acbd}u^c u^d\,,~~~~ H_{ab}={\ts{1\over2}}\ep_{acd}
C^{cd}{}{}_{be}u^e \,,
$$
where $\,h_{ab}=g_{ab}+u_au_b\,$ is the spatial projector, 
$\,g_{ab}\,$ is the metric tensor, and
$\,\ep_{abc}=\eta_{abcd}u^d\,$ is the spatial projection of the
spacetime permutation tensor $\,\eta_{abcd}\,$ \cite{m}. 
Gravitational radiation is covariantly described by the nonlocal
fields $\,E_{ab}\,$, the tidal part of the curvature which
generalizes the Newtonian tidal tensor, and $\,H_{ab}\,$, which has
no Newtonian analogue \cite{ev}. As such, $\,H_{ab}\,$ may be 
considered as the true gravity wave tensor, since there is no
gravitational radiation in Newtonian theory. However, as in 
electromagnetic theory, gravity waves are characterized by 
$\,H_{ab}\,$ and $\,E_{ab}\,$, where both are divergence--free
but neither is curl--free \cite{he}, \cite{dbe}.

In \cite{m}, it was shown that in the generic case, 
i.e., without imposing any divergence--free conditions,
the covariant constraint equations 
evolve consistently with
the covariant propagation equations. These equations are:\\
{\em Propagation equations}
\begin{eqnarray}
\dot{\rho}+\Theta\rho &=& 0\,, 
\label{p1}\\
\dot{\Theta}+{\ts{1\over3}}\Theta^2 +{\ts{1\over2}}\rho &=&
-\sigma_{ab}\sigma^{ab}\,,
\label{p2}\\
\dot{\sigma}_{ab}+{\ts{2\over3}}\Theta\sigma_{ab}+E_{ab} &=&
-\sigma_{c\langle a}\sigma_{b\rangle }{}^c \,,
\label{p3}\\
\dot{E}_{ab}+\Theta E_{ab}
-\c H_{ab}+{\ts{1\over2}}\rho\sigma_{ab}  &=&
3\sigma_{c\langle a}E_{b\rangle }{}^c \,,
\label{p4}\\
\dot{H}_{ab}+\Theta H_{ab}+\c E_{ab} &=&
3\sigma_{c\langle a}H_{b\rangle }{}^c \,.
\label{p5}
\end{eqnarray}
{\em Constraint equations}
\begin{eqnarray}
\d^b\sigma_{ab} - {\ts{2\over3}}\d_a \Theta &=&0\,,
\label{c1}\\
\c \sigma_{ab}- H_{ab} &=&0\,,
\label{c2}\\
\d^b E_{ab} - {\ts{1\over3}}\d_a \rho &=&
\ep_{abc}\sigma^b{}_d H^{cd}\,,
\label{c3}\\
\d^b H_{ab} &=& -\ep_{abc}\sigma^b{}_d E^{cd}\,,
\label{c4}
\end{eqnarray}
where
$\,S_{\langle ab\rangle }=h_a{}^c h_b{}^d S_{(cd)}-
{\ts{1\over3}}S_{cd}h^{cd} h_{ab}\,$ is the projected, symmetric
and trace--free part of $\,S_{ab}\,$, the covariant 
spatial derivative is defined by
$\,\d_a S^{b\cdots}{}{}_{\cdots c}=h_a{}^p h^b{}_q \cdots
h_c{}^r \nabla_p
S^{q\cdots }{}{}_{\cdots r}\,$, 
the covariant spatial
divergence is $\,\d^bS_{ab}\,$, and the covariant spatial
curl is $\,\c S_a=\ep_{abc}\d^bS^c\,$ for vectors and
$\,\c S_{ab}=\ep_{cd(a}\d^c S_{b)}{}^d\,$
for tensors. (Further details are given
in \cite{m}, \cite{mes}.) 
In the linearized theory of covariant perturbations about an
FRW background, the right sides of these equations are all zero.

It was previously claimed that in the exact nonlinear theory, the
gravity wave condition
\begin{equation}
\d^bH_{ab}=0
\label{dh}\end{equation}
implies $\,H_{ab}=0\,$ \cite{led}. As shown in \cite{m}, this claim
arises from a sign error and is incorrect, Bianchi type V spacetimes
providing a counterexample.
Here we complete the analysis of \cite{m} by showing 
that consistency is maintained if (\ref{dh}) is imposed, 
without $\,H_{ab}\,$ zero.

The fact that consistency is not automatic is illustrated by
the case of silent universes, in which $\,H_{ab}=0\,$. For
these solutions, 
consistent evolution of the condition $\,H_{ab}=0\,$ 
imposes a
series of nontrivial integrability conditions, which are
identically satisfied in the linearized case, but not
in the nonlinear case \cite{m2}, \cite{vulem}. Thus there is
a linearization instability in silent universes. By contrast,
when (\ref{dh}) holds but $\,H_{ab}\,$ is not forced to vanish, 
which includes 
gravity wave solutions, there is no linearization instability
following from the evolution of (\ref{dh}).
An example of consistency conditions arising already at the linearized
level is given by purely magnetic spacetimes, $\,E_{ab}=0\,$, 
for which
$\,\Theta\sigma_{ab}=0\,$ \cite{m2}.

The proof that (\ref{dh}) evolves consistently is based on
a combination of tetrad methods \cite{led}, \cite{l} and
the covariant methods of \cite{m}. 
The only direct effect of (\ref{dh}) on the covariant propagation
and constraint equations
is an algebraic modification of the constraint (\ref{c4}), which
does not change the consistent evolution of the constraints.
We have to check only consistent evolution
of the new condition
(\ref{dh}) itself. It is more convenient to replace (\ref{dh}) by
the equivalent condition that follows from (\ref{c4}),
\begin{equation}
[\sigma,E]=0\,,
\label{b}\end{equation}
where we are using index--free notation
for the covariant commutator. In the linearized case,
(\ref{b}) is identically 
satisfied since the left side is second order of smallness, and
consistency is automatic.

In the exact nonlinear case,
using only the shear propagation equation (\ref{p3}) and its
covariant time derivative,
we find that
$$
[\sigma,\dot{E}] = -[\sigma,\ddot{\sigma}]+{\ts{2\over3}}\Theta
[\sigma,E]-\sigma[\sigma,E]
$$
and
$$
[\dot{\sigma},E] = -{\ts{2\over3}}\Theta[\sigma,E]+\sigma
[\sigma,E]\,.
$$
Adding these equations gives
\begin{equation}
[\sigma,E]^{\rd}=-[\sigma,\ddot{\sigma}]\,.
\label{bdot}\end{equation}
Now the 
right side may be shown to vanish identically
without differentiating (\ref{b}), i.e. using only the algebraic
content of (\ref{b}), as follows. 

From the shear propagation equation (\ref{p3}), (\ref{b}) is
equivalent to
\begin{equation}
[\sigma,\dot{\sigma}]=0\,.
\label{b2}\end{equation}
We choose an orthonormal tetrad \cite{e} $\,\{{\bf e}_0={\bf u}\,,
{\bf e}_\mu\}\,$, with $\,\{{\bf e}_\mu\}\,$ a shear eigenframe, so 
that
\begin{equation}
\sigma_{0a}=0=\partial_0\sigma_{0a}\,,
~~\sigma_{\mu\nu}=0=\partial_0\sigma_{\mu\nu}~~\mbox{if}~~
\mu\neq\nu\,,
\label{s}\end{equation}
where $\,\partial_0\,$ denotes the directional derivative 
along $\,{\bf e}_0={\bf u}\,$. 
Then we have
\begin{equation}
[\sigma,\dot{\sigma}]_{ab}=\left(\sigma_{aa}-\sigma_{bb}\right)
\dot{\sigma}_{ab}~~~(\mbox{no sum})\,.
\label{s2}\end{equation}
At all points where the shear is nondegenerate, i.e., where
$\,\sigma_{aa}\neq\sigma_{bb}\,$ when $\,a\neq b\,$,
(\ref{s2}) and (\ref{b2}) 
show that $\,\dot{\sigma}_{ab}\,$ is diagonal -- and 
thus $\,E_{ab}\,$ is also diagonal, by (\ref{p3}). 
In fact diagonality still
holds at points of degeneracy, as follows from the tetrad
form of the covariant derivative:
$$
\dot{\sigma}_{ab}=\partial_0\sigma_{ab}-\Gamma^c{}_{0b}\sigma_{ac}
-\Gamma^c{}_{0a}\sigma_{cb}\,,
$$
where the Ricci rotation coefficients are $\,\Gamma_{abc}=
{\bf e}_a\cdot\nabla_b {\bf e}_c=-\Gamma_{cba}\,$.
Using (\ref{s}) and (\ref{p3}), we get
\begin{equation}
a\neq b~~\Rightarrow~~\dot{\sigma}_{ab}=\left(\sigma_{aa}-
\sigma_{bb}\right)\Gamma_{b0a}=-E_{ab}~~(\mbox{no sum})\,,
\label{s3}\end{equation}
so that $\dot{\sigma}_{ab}\,$ is diagonal also where
$\,\sigma_{aa}=\sigma_{bb}\,$ ($\,a\neq b\,$).
Thus {\em the shear eigenframe simultaneously 
diagonalizes $\,\sigma_{ab}\,$,
$\,\dot{\sigma}_{ab}\,$ and $\,E_{ab}\,$}. 
This regains a result given in \cite{br}.

It also follows from (\ref{b2}), (\ref{s2}) and (\ref{s3}) that
\begin{equation}
\Gamma_{a0b}=0 
\label{r}\end{equation}
holds at all points where the shear is nondegenerate. 
(Note that (\ref{r})
is an identity for $\,a=b\,$.) 
At points of degeneracy, i.e., where
$\,\sigma_{11}=\sigma_{22}\,$, we can use the remaining tetrad
freedom of a rotation in the $\,\{{\bf e}_1,{\bf e}_2\}\,$ plane
to set $\,\Gamma_{102}=0\,$, so that (\ref{r}) 
still holds. Specifically, such a rotation through an
angle $\,\alpha\,$ 
preserves (\ref{s}) and the degeneracy, while
$$
\Gamma_{102} ~~\rightarrow ~~ \Gamma_{102}-\partial_0\alpha \,.
$$
Thus we can ensure that (\ref{r}) holds throughout spacetime,
by specializing the eigenframe where necessary.
Then (\ref{r}) shows that $\,\ddot{\sigma}_{ab}\,$ is
also diagonal in this frame, since
$$
a\neq b~\Rightarrow~
\ddot{\sigma}_{ab}=(\dot{\sigma}_{aa}-\dot\sigma_{bb})
\Gamma_{b0a}=0~~(\mbox{no sum})\,,
$$
where we have used the fact that $\,\partial_0\dot{\sigma}_{ab}\,$
is diagonal.
The covariant (frame--independent)
consequence of the simultaneous diagonalizability of
$\,\sigma_{ab}\,$ and $\,\ddot{\sigma}_{ab}\,$ is
$$
[\sigma,\ddot{\sigma}]=0\,,
$$
which shows that the right side of (\ref{bdot}) does indeed 
vanish identically,
consistent with and
independent of the derivative of (\ref{b}). 
Thus the
the first covariant time derivative of the condition (\ref{b})
imposes no consistency conditions. It is clear from the above 
argument that all the subsequent covariant time derivatives
of $\,\sigma_{ab}\,$ are also diagonal in the eigenframe, so that
these higher derivatives all commute with the shear and amongst
themselves. It follows that the second and higher covariant time
derivatives of the condition (\ref{b}) also vanish without
further conditions. 

This establishes that {\em the covariant condition}
$\,\div H=0\,$ {\em evolves consistently
in the exact nonlinear theory.}
The question whether such consistency extends to the further covariant
gravity wave condition
$\,\div E=0\,$ is more difficult, and under investigation.

Finally, we note that, by virtue of (\ref{r}) and the propagation
equation (\ref{p4}), $\,\c H_{ab}\,$ is also diagonal in the 
eigenframe that diagonalizes $\,\sigma_{ab}\,$ and $\,E_{ab}\,$,
i.e., {\em there is a shear eigenframe such that} $\,\sigma_{ab}\,$,
$\,E_{ab}\,$, $\,\c H_{ab}\,$ 
{\em and all their covariant time
derivatives are diagonal, and therefore commute.}

\end{document}